\documentclass[preprint,aps,nofootinbib,showpacs,amsfonts,epsf]{revtex4}

\usepackage{amssymb}
\input{epsf.tex}

\newcommand{\Pint}{\diagup\hspace{-0.45cm}\int}
\newcommand{\E}{{\cal{E}}}

\newcommand{\s}{\sigma}

\renewcommand{\a}{\alpha}

\newcommand{\be}{\begin{equation}}
\newcommand{\ee}{\end{equation}}
\newcommand{\bea}{\begin{eqnarray}}
\newcommand{\eea}{\end{eqnarray}}
\newcommand{\ba}{\begin{array}}
\newcommand{\ea}{\end{array}}

\begin{document}
\draft

\title{On interrelations between Sibgatullin's and Alekseev's\\ approaches to the construction of exact solutions\\ of the Einstein-Maxwell equations}

\author{F J Ernst$^1$, V S Manko$^2$ and E Ruiz$^3$}
\address{$^1$FJE Enterprises, 511 CR 59, Potsdam, NY 13676, USA\\
$^2$Departamento de F\'\i sica, Centro de Investigaci\'on y de
Estudios Avanzados del IPN, A.P. 14-740, 07000 M\'exico D.F., Mexico\\
$^3$Instituto Universitario de F\'{i}sica
Fundamental y Matem\'aticas, Universidad de Salamanca, 37008 Salamanca, Spain}

\begin{abstract}
The integral equations involved in Alekseev's ``monodromy transform'' technique are shown to be simple combinations of Sibgatullin's integral equations and normalizing conditions. An additional complex conjugation introduced by Alekseev in the integrands makes his scheme mathematically inconsistent; besides, in the electrovac case all Alekseev's principal value integrals contain an intrinsic error which has never been identified before. We also explain how operates a non--trivial double--step algorithm devised by Alekseev for rewriting, by purely algebraic manipulations and in a different (more complicated) parameter set, any particular specialization of the known analytically extended $N$--soliton electrovac solution obtained in 1995 with the aid of Sibgatullin's method. \end{abstract}

\pacs{04.20.Jb}

\maketitle


\section{Introduction}

The integral equations and normalizing conditions involved in Sibgatullin's method have the form \cite{Sib}
\bea &&\Pint_{-1}^{+1}\frac{e(\xi)+\tilde
e(\eta)+2f(\xi)\tilde f(\eta)}
{(\xi-\eta)\sqrt{1-\s^2}}\mu_a(\s)d\s=-2\pi i(\delta_a^2-2\tilde
f(\eta)\delta_a^3), \label{eqs_S} \\
&&\int_{-1}^{+1}\frac{\mu_a(\s)d\s}
{\sqrt{1-\s^2}}=\pi\delta_a^1, \quad a=1,2,3, \label{cond_n} \eea
and they permit one to calculate the unknown functions $\mu_a(\s)$ corresponding to the arbitrarily prescribed data $e(\xi)$ and $f(\xi)$, $\xi=z+i\rho\s$, representing local holomorphic continuations of the axis expressions $e(z)\equiv\E(\rho=0,z)$, $f(z)\equiv\Phi(\rho=0,z)$ of the complex Ernst potentials \cite{Ern} into the complex plane. The knowledge of $\mu_a(\s)$ is essential for being able to subsequently obtain $\E(\rho,z)$, $\Phi(\rho,z)$ and metric coefficients by virtue of integral formulae (see \cite{MSi} for details).

Whereas Sibgatullin's method is based on the Hauser--Ernst formulation of Kinnersley's transformation theory \cite{HEr,Kin1} and all of its details are well known, the so--called ``monodromy transform'' technique first presented by Alekseev under a different name and in an axiomatic manner \cite{Ale} has been always advertised by its author as an independent mathematical result and an alternative to the Hauser, Ernst and Sibgatullin's use of the homogeneous Riemann--Hilbert problem in their solution generating methods. In the stationary axially symmetric electrovac case, Alekseev's integral equations for the unknown functions $\varphi^a(\zeta)$ have the form
\be
\frac{1}{\pi i}\Pint_L\frac{{\cal K}(\tau,\zeta)}{\zeta-\tau}
\varphi^a(\zeta)d\zeta=k^a(\tau), \quad a=1,2,3, \label{eqs_A1} \ee where \bea {\cal
K}(\tau,\zeta)&=&-[\lambda]_\zeta\Bigl(\vec k(\tau)\cdot\vec
l(\zeta)\Bigr)=-[\lambda]_\zeta\Bigl(
1+i\zeta(u^\dag(\zeta)-u(\tau))
+4\zeta^2v^\dag(\zeta)v(\tau)\Bigr), \nonumber\\ \vec
k(\tau)&=&\{ 1,u(\tau),v(\tau)\}, \quad \vec l(\zeta)=\left(
\begin{array}{c}
1+i\zeta u^\dag(\zeta)\\ -i\zeta\\
4\zeta^2v^\dag(\zeta)
\end{array}
\right). \label{eqs_A2} \eea

In our short communication we will show that the above equations (\ref{eqs_A1}) and (\ref{eqs_A2}) are simple combinations of Sibgatullin's equations (\ref{eqs_S}) and (\ref{cond_n}). We will also describe the algebraic trick invented by Alekseev for presenting the already known solutions obtained via Sibgatullin's method as an independent product of the ``monodromy transform'' procedure.

\section{Derivation of Alekseev's equations}

We first observe that Alekseev's $\zeta$, $\tau$ and the `$\dag$' operation are identical, respectively, to Sibgatullin's $\xi$, $\eta$ and the `$\sim$' operation ($\tilde g(\xi)\equiv\overline{g(\bar\xi)}$, where a bar over a symbol means complex conjugation). The integration in (\ref{eqs_A1}) along the cut $L$ joining the points $z\pm i\rho$ reduces to the integration in (\ref{eqs_S}) by parametrizing the points of the cut as $\zeta=z+i\rho\s$, $\s\in[-1,1]$, and taking into account that $d\zeta=i\rho d\s$.

Since in the electrovac case Alekseev's integral equations (\ref{eqs_A1}) and (\ref{eqs_A2}) are erroneous, we find it instructive to restrict our consideration here to only the pure vacuum case and after that briefly comment on the mistakes in Alekseev's formulae involving the function $v(\zeta)$. In the absence of the electromagnetic field ($v(\zeta)=0$ in (\ref{eqs_A2}) and $f(\xi)=0$ in (\ref{eqs_S})) we have two Alekseev's equations
\bea &&\frac{1}{\pi
i}\Pint_L\frac{[\lambda]_\zeta\Bigl(1+i\zeta(u^\dag
(\zeta)-u(\tau))\Bigr)} {\zeta-\tau} \varphi^1(\zeta)d\zeta=-1,
\nonumber\\ &&\frac{1}{\pi
i}\Pint_L\frac{[\lambda]_\zeta\Bigl(1+i\zeta(u^\dag
(\zeta)-u(\tau))\Bigr)} {\zeta-\tau}
\varphi^2(\zeta)d\zeta=-u(\tau), \label{eqs_Av} \eea
and two Sibgatullin's integral equations
\be
\frac{1}{2\pi i}\Pint_{-1}^{+1}\frac{\tilde e(\xi)+e(\eta)}
{(\xi-\eta)\sqrt{1-\s^2}}\mu_1(\s)d\s=0, \quad
\frac{1}{2\pi i}\Pint_{-1}^{+1}\frac{\tilde e(\xi)+e(\eta)}
{(\xi-\eta)\sqrt{1-\s^2}}\mu_2(\s)d\s=1, \label{eqs_Sv} \ee with a pair of normalizing
conditions for $\mu_1(\s)$ and $\mu_2(\s)$:
\be
\int_{-1}^{+1}\frac{\mu_1(\s)d\s} {\sqrt{1-\s^2}}=\pi, \quad
\int_{-1}^{+1}\frac{\mu_2(\s)d\s} {\sqrt{1-\s^2}}=0. \label{cond_nv}\ee
Note that for having full coincidence with Alekseev's expressions we have made in (\ref{eqs_Sv}) the additional `tilde' conjugation which means that in the initial data we have conjugated the complex parameters.

Let us now choose $e(\xi)$ and $e(\eta)$ in the form
\be
e(\xi)=1-2i\xi u(\xi), \quad e(\eta)=1-2i\eta u(\eta), \quad
\tilde e(\xi)=1+2i\xi\tilde u(\xi).\ee Then we have \be \tilde
e(\xi)+e(\eta)=2\{1+i\xi(\tilde
u(\xi)-u(\eta))+i(\xi-\eta)u(\eta)\}. \label{kern1} \ee

Substituting (\ref{kern1}) into equations (\ref{eqs_Sv}) and taking into account that
\be
\int_{-1}^{+1}\frac{i u(\eta)\mu_1(\s)d\s} {\sqrt{1-\s^2}}=\pi i
u(\eta) \quad \hbox{and} \quad \int_{-1}^{+1}\frac{u(\eta)\mu_2(\s)d\s} {\sqrt{1-\s^2}}=0 \label{int} \ee
by virtue of (\ref{cond_nv}), we finally get
\be
\frac{1}{\pi i}\Pint_{-1}^{+1}\frac{1+i\xi(\tilde
u(\xi)-u(\eta))} {(\xi-\eta)\sqrt{1-\s^2}}\mu_1(\s)d\s=-u(\eta), \quad \frac{1}{\pi i}\Pint_{-1}^{+1}\frac{1+i\xi(\tilde
u(\xi)-u(\eta))} {(\xi-\eta)\sqrt{1-\s^2}}\mu_2(\s)d\s=1, \label{eqs_fin} \ee
which are precisely Alekseev's equations (\ref{eqs_Av}), up to obvious redefinitions of the functions $\varphi^1$ and $\varphi^2$, and the change of the integration variable (these are left to the reader as a simple exercise).

Mention that Alekseev's manner of combining Sibgatullin's integral equations and normalizing conditions can hardly be recommended for practical applications as it complicates the technical part of the solution construction procedure. It should also be emphasized that due to the additional `tilde' operation performed by Alekseev in the integral formulae his scheme turns out {\it mathematically inconsistent} because the final expressions for the Ernst potentials obtained in this way do not reduce to the initial axis data, but to the complex conjugate quantities only.

In the electrovac case ($f(\xi)\ne0$, $f(\xi)=2i\xi v(\xi)$) we have
\be \tilde e(\xi)+e(\eta)+2\tilde f(\xi)f(\eta)
=2\{1+i\xi(\tilde u(\xi)-u(\eta))+4\xi\eta\tilde v(\xi)v(\eta)
+i(\xi-\eta)u(\eta)\}, \ee
and Sibgatullin's integral equation for the function $\mu_1(\s)$, in which the additional `tilde' conjugation is performed and the respective normalizing condition is used in analogy with the vacuum case, then rewrites as
\be
\frac{1}{\pi i}\Pint_{-1}^{+1}\frac{1+i\xi(\tilde
u(\xi)-u(\eta))+4\xi\eta\tilde v(\xi)v(\eta)}
{(\xi-\eta)\sqrt{1-\s^2}}\mu_1(\s)d\s=-u(\eta). \label{eq1_S} \ee
This means that the kernel ${\cal K}(\tau,\zeta)$ of all three Alekseev's equations (\ref{eqs_A1}) contains an intrinsic error -- the factor $\zeta^2$ -- whose changing to the correct factor $\zeta\tau$ also implies modifications of $\vec k(\tau)$ and $\vec l(\zeta)$ in (\ref{eqs_A2}).

\section{The algebraic trick}

The use of Sibgatullin's method is particularly simple in the case of the rational axis data. In 1995 the extended $N$--soliton electrovac solution was constructed with its help in a concise analytical form \cite{RMM}, the solution arising from the axis data
\bea
e(z)=1+\sum\limits_{l=1}^N\frac{e_l}{z-\beta_l}, \quad
f(z)=\sum\limits_{l=1}^N\frac{f_l}{z-\beta_l}, \label{ad_Ns}
\eea
where $e_l$, $\beta_l$ and $f_l$ are $3N$ arbitrary complex constants. Two particular ($N=1$ and $N=2$) electrovac specializations of this solution were considered in a recent paper \cite{ABe} within the framework of the ``monodromy transform'' technique. A careful inspection of the paper \cite{ABe} reveals, however, that the solution generating procedure employed in \cite{ABe} reduces exclusively to rewriting in different parameters the already known results  using the following simple algebraic trick devised by Alekseev for avoiding the use of his problematic integral equations. Alekseev starts with the unphysical, asymptotically non--flat axis data
\be
e(z)=1-2i\sum\limits_{l=1}^N u_l-2i\sum\limits_{l=1}^N\frac{u_lh_l}{z-h_l}, \quad f(z)=2i\sum\limits_{l=1}^N v_l+2i\sum\limits_{l=1}^N\frac{v_lh_l}{z-h_l} \label{data1_A} \ee
($u_l$, $h_l$ and $v_l$ are complex parameters) and {\it formally} declares that for those data he has carried out the solution construction procedure with the aid of the `monodromy transform' technique (still having nothing in the hand) and that for writing down the final result he needs to apply to the `obtained solution' an additional `gauge transformation' of the form
\bea\E'&=&1+\frac{\E-\varepsilon_0+2\bar\phi_0(\Phi-\phi_0)}{p_0}, \quad \Phi'=\frac{\Phi-\phi_0}{\sqrt{p_0}}, \\ \varepsilon_0&\equiv&1-2i\sum\limits_{l=1}^N u_l, \quad \phi_0\equiv 2i\sum\limits_{l=1}^N v_l, \quad \bar\phi_0\equiv -2i\sum\limits_{l=1}^N\bar v_l, \quad p_0\equiv 1-i\sum\limits_{l=1}^N(u_l-\bar u_l)+\phi_0\bar\phi_0. \nonumber \label{gt_g} \eea
The only purpose of the above transformation is to cast the unphysical axis data (\ref{data1_A}) to the new data $e'(z)$ and $f'(z)$ identical to (\ref{ad_Ns}) but written in a different parameter set:
\be
e'(z)=1+\sum\limits_{l=1}^N\frac{2ih_l(-u_l+2\bar\phi_0v_l)}{p_0(z-h_l)}, \quad f'(z)=\sum\limits_{l=1}^N\frac{2iv_lh_l}{\sqrt{p_0}(z-h_l)}, \label{data2_A} \ee
and these expressions provide Alekseev with the desired relations between his parameters $u_l$, $h_l$, $v_l$ and the parameters $e_l$, $\beta_l$, $f_l$ entering the axis data (\ref{ad_Ns}) of the known analytically extended $N$--soliton solution:
\be
e_l=\frac{2ih_l}{p_0}(-u_l+2\bar\phi_0v_l), \quad \beta_l=h_l, \quad f_l=\frac{2iv_lh_l}{\sqrt{p_0}}. \label{elul} \ee

The inverted formulae are
\be u_l=\frac{i p_0}{2\beta_l}\left(e_l-2f_l\sum\limits_{k=1}^N\frac{\bar f_k}{\bar \beta_k}\right), \quad h_l=\beta_l, \quad v_l=-\frac{if_l\sqrt{p_0}}{2\beta_l}, \quad p_0\equiv\frac{\prod_{l=1}^N\beta_l\bar\beta_l} {\prod_{n=1}^{2N}\a_n}, \label{ulel}    \ee
where $\a_n$ are $2N$ formal roots of the algebraic equation $e(\xi)+\tilde e(\xi)+2f(\xi)\tilde f(\xi)=0$.

Since the $N$--soliton electrovac solution in terms of the parameters $e_l$, $\beta_l$, $f_l$ is known, the substitution of (\ref{elul}) into the formulae of the paper \cite{RMM} immediately supplies Alekseev with the explicit form of the multisoliton solution written in terms of the parameters $u_l$, $h_l$, $v_l$. Apparently, the algebraic trick described above, if not known, causes an illusion of a true usage of the integral equations in obtaining Alekseev's results and novelty of the latter. We point out that writing the soliton solution in terms of the parameter set $\{u_l, h_l, v_l\}$ instead of the canonical set $\{e_l, \beta_l, f_l\}$ considerably complicates the solution's form and hence does not have any scientific justification.

In section~2 of \cite{ABe} the above formulae (\ref{elul}) and (\ref{ulel}) are used for reproducing the results of the paper \cite{AGM} devoted to the derivation of the physical representation of the electrovac Demia\'nski--Newman solution \cite{DNe}. Just to have an idea about the advantages of the original work (not even cited in \cite{ABe}) over its remake, it suffices to say that the axis data employed in \cite{AGM} have the form
\be e(z)=\frac{z-m-i(a+\nu)}{z+m-i(a-\nu)}, \quad f(z)=\frac{q+i b}{z+m-i(a-\nu)}, \label{ad_DN} \ee
while the initial expression for $e(z)$ in terms of the same parameters considered in \cite{ABe} is the following complicated function:
\be
e(z)=-1-\frac{2ia[m-i(a-\nu)]}{m^2+\nu^2-a^2-q^2-b^2}+\frac{2[m-i(a-\nu)]} {z+m-i(a-\nu)}\left(1+\frac{ia[m-i(a-\nu)]}{m^2+\nu^2-a^2-q^2-b^2}\right). \label{ez_DNA} \ee

Turning now to the two--pole electrostatic solution considered in section~3 of \cite{ABe}, it should be first of all pointed out that the solution defined by formulae (34)--(38) of that paper, contrary to the authors' claim, {\it does not correspond} to the axis data (20), (21) of \cite{ABe}. The algebraic trick is used this time for rewriting the Bret\'on {\it et al} double--Reissner--Nordstr\"om solution \cite{BMA} obtained in 1998 in a closed analytical form involving the canonical parameter set. Formulae (20)--(29) of \cite{ABe} utilize the relations
\be u_l=\frac{i p_0}{2\beta_l}\left(e_l-2f_l\sum\limits_{k=1}^2\frac{f_k}{\beta_k}\right), \quad h_l=\beta_l, \quad v_l=-\frac{if_l\sqrt{p_0}}{2\beta_l}, \quad p_0\equiv\frac{\beta_1^2\beta_2^2} {\a_1\a_2\a_3\a_4}, \label{vlfl}  \ee
for obtaining the form of $u_l$ and $v_l$ in terms of $h_l$, $w_l$, $\tilde w_l$ (i.e., $\beta_l$, $\a_n$ in the notions of \cite{BMA}) by substituting into (\ref{vlfl}) the explicit expressions for $e_l$ and $f_l$ derived in \cite{BMA}. It is worth mentioning in conclusion that formulae (34)--(38) of \cite{ABe} describe a 5--parameter representation of the BMA solution in physical parameters first obtained in \cite{Man} and then just rewritten in \cite{ABe} in different coordinates.

\section*{Acknowledgments}

This work was partially supported by Project FIS2006-05319 from Ministerio de Ciencia y Tecnolog\'\i a, Spain, and by the Junta de Castilla y Le\'on under the ``Programa de Financiaci\'on de la Actividad Investigadora del Grupo de Excelencia GR-234'', Spain.

\end{document}